\documentclass[%
 reprint,
superscriptaddress,
 amsmath,amssymb,
 aps,
]{revtex4-2}
\usepackage{subfigure}
\usepackage{graphicx}
\usepackage{dcolumn}
\usepackage{bm}

\begin{document}

\preprint{APS/123-QED}

\title{A First-Principle Approach to X-ray Active Optics: Design and  Verification}

\author{Dezhi Diao}
 \affiliation{Institute of High Energy Physics, Chinese Academy of Sciences, Beijing 100049, China}
 \affiliation{University of Chinese Academy of Sciences, Beijing 100049, China}
 
\author{Han Dong}
 \affiliation{Institute of High Energy Physics, Chinese Academy of Sciences, Beijing 100049, China}
 \affiliation{University of Chinese Academy of Sciences, Beijing 100049, China}
 
\author{Fugui Yang }
 \email{yangfg@ihep.ac.cn}
 \affiliation{Institute of High Energy Physics, Chinese Academy of Sciences, Beijing 100049, China}
 
\author{Ming Li}
 \affiliation{Institute of High Energy Physics, Chinese Academy of Sciences, Beijing 100049, China}

\author{Weifan Sheng}
 \affiliation{Institute of High Energy Physics, Chinese Academy of Sciences, Beijing 100049, China}

\author{Xiaowei Zhang }
 \email{zhangxw@ihep.ac.cn} 
 \affiliation{Institute of High Energy Physics, Chinese Academy of Sciences, Beijing 100049, China}

\date{\today}

\begin{abstract}
This paper presents the first-principle design approach for X-ray active optics, using the simulation-modulation cycle in place of the measurement-modulation feedback loops used in traditional active optics. Hence, the new active optics have the potential to outperform the accuracy of surface-shape metrology instruments. We apply an X-ray mirror with localized thermal elastic deformation to validate the idea. Both the finite element simulations and surface shape measurements have demonstrated that the active optics modulation accuracy limit can be achieved at the atomic layer level. It is believed that the implementation of the first-principle design strategy has the capacity to revolutionize both the manufacturing processes of X-ray mirrors and the beamline engineering of synchrotron radiation.
\end{abstract}

\maketitle

At the cutting edge of engineering and technology, obtaining large-size, ultra-high-accuracy mirrors has proven to be a recurring difficulty \cite{graves2019precision,moller2019wafer}. This involves ultra-precision machining and measurement techniques. Nevertheless, these optics often need to operate under extreme conditions, where factors such as thermal deformations induced by the beam and deformations due to mirror clamping surpass the precision of mirror manufacturing. Simply striving for accuracy during production is no longer adequate to fulfill realistic usage demands. Therefore, the introduction of new technological elements like active optics becomes imperative to address these challenges. Successful practices on large-scale ultra-precision optical instruments such as LIGO and Telescope \cite{tyson1999adaptive}, along with advances in traditional optics and optical engineering, demonstrate that active optics is a fundamental strategy in system construction.

Synchrotron radiation facilities and their x-ray mirrors are categorized as large-scale precision equipment. The low grazing incident angle of synchrotron X-ray mirrors poses challenges in achieving precise surface shape during both fabrication and operation in the beamline. However, the elongated dimension of X-ray mirror in tangential direction provide ample space for mirror surface modulations, and the one-dimensional adjustment requirement makes the approach to X-ray mirror surface shape correction relatively straightforward. To make X-ray active optics work better, different methods have been developed since the 1990s. These include piezoelectric \cite{susini1995adaptive, alcock2023fast}, mechanically bent \cite{howells2000theory,goto2018nearly}, and heating technologies \cite{cocco2020adaptive}. The primary constituents of active optics systems consist of deformable mirror mechanics and metrology. Due to the challenges in high-precision surface shape metrology and the non-linear response of deformable mirrors to the drive, the traditional operation mode requires iterative procedures which includes measuring surface shape, adjusting deformable mirrors based on feedback, and ultimately achieving convergence. Hence, the level of accuracy achieved in active optics is dependent upon the precision of surface shape measurements and the intrinsic precision of the actuator. Compared to metrology in the manufacture of X-ray mirrors, this in-situ high-precision surface-shape metrology technique presents more significant challenges. The need for high-precision devices, including actuators and metrology equipment, unavoidably contributes to the total cost. Hence, it is pragmatically crucial to develop robust, efficient, and cost-effective methodologies and technologies for X-ray active optics.
\begin{figure*}
\includegraphics[width=0.7\textwidth]{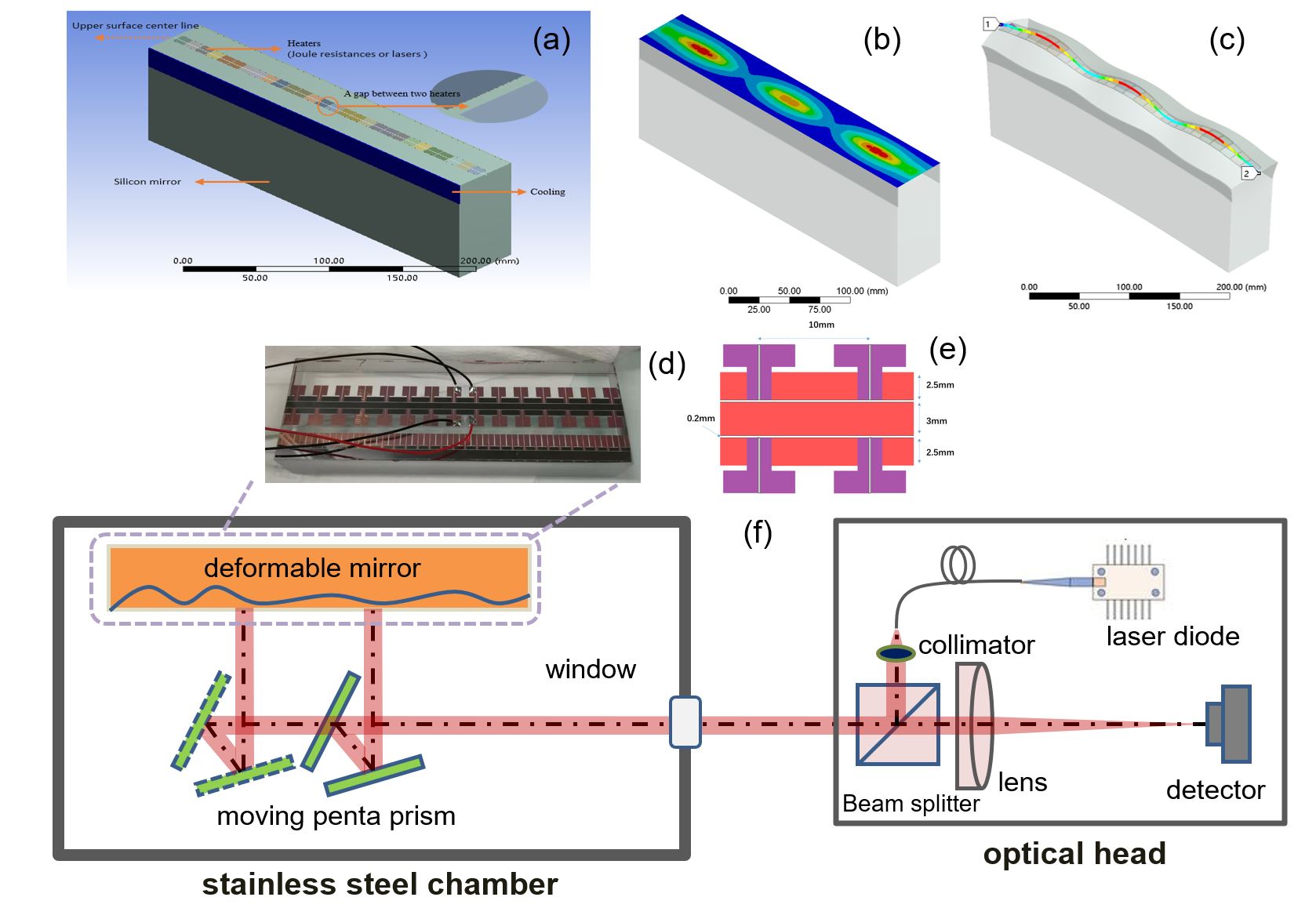}
\caption{\label{fig:wide}The first-principles-designed X-ray active optics system. (a) the model of an X-ray mirror with thermal deformation modulation; (b) and (c) depict the temperature field distribution and thermo-elastic deformation calculated through FEA, respectively. A demon's active mirror (d) is divided into different areas for heating and measurement with details in (e). (f) is measurement setup of the surface shape modulation test experiment with long trace profiler.}
\end{figure*}

In this paper, we introduce and explore the application of a first-principles approach for designing X-ray active optics. The strategy involves simplifying technical elements within the system and ensuring deterministic control as a fundamental assurance for achieving the target precision with low risk. To achieve deterministic control, a design rooted in fundamental principles is crucial, encompassing the careful selection of controllable physical processes and the utilization of appropriate engineering implementation methods. In addition, significant progress in computer science has enhanced human capabilities and provided solutions for complex engineering challenges, enabling the implementation of first-principles design. In contrast to the normative "copying principle," which prioritizes precision in all technical elements and aspects, we believe that the computational-assisted first-principles design approach not only propels technological advancement but also significantly reduces costs and development cycles.

To realize the concepts of first-principles design, we have developed a system for X-ray active optics that utilizes the thermo-elastic deformation of the x-ray mirror, namely the thermal-deformable mirror.  The modulation system in concern employs an array of space-distributed thermal sources, such as laser heating, electrical heating, or cooling-heating, to stimulate localized thermal deformation on the mirror surface. This process effectively modulates the one-dimensional shape of the X-ray mirror surface. The system implementation is illustrated in Fig. 1.  Fig. 1 (a) shows the configuration of heating devices on the mirror surface. These components are strategically placed next to or partially covering the area of the X-ray reflecting mirror to enhance the modulation performance. "First-principles design" is also the accurate prediction of the necessary configuration of heating elements and amounts of deformation using thermal elasticity theory. The general heat conduction equation for a temperature distribution $T$ within a body has the form \cite{hahn2012heat}:
\begin{equation}
\nabla^2 T(\mathbf{r}, t)=-\frac{1}{k} g(\mathbf{r}, t)%
\end{equation}

where $\alpha=k / \rho c_p$ is the thermal diffusivity and $g(\mathbf{r}, t)$ is the heat generation or source term.  If we call $\vec{u}(r, z)$ the displacement vector, i.e., the difference between the coordinates of a given atom before and after heating, the classically relevant quantities are the strain tensor $\varepsilon_{i j}$ and the stress tensor $\theta_{i j}$. In the presence of a temperature field $T$, the two tensors are related by the generalized Hooke law for isotropic media via the Lamé coefficients $\lambda, \mu$:
\begin{equation}
\theta_{i j}=\delta_{i j}(\lambda \varepsilon-\nu T)+2 \mu \varepsilon_{i j}%
\end{equation}
where $\nu$ is the stress temperature modulus and $\varepsilon$ is the trace of the strain tensor. $\delta_{i j}$ is the Kronecker tensor. The strain tensor $\varepsilon_{i j}$ is related to the displacement field $\vec{u}(r, z)$ through the displacement gradient $(\nabla \vec{u})$ as follows:
\begin{equation}
\varepsilon_{i j}=\frac{1}{2}\left(\frac{\partial u_i}{\partial x_j}+\frac{\partial u_j}{\partial x_i}\right)
\end{equation}
These equations, when appropriately coupled and solved along with suitable initial and boundary conditions, can describe the evolution of temperature and displacement fields in an object undergoing elastic deformation due to heating. We solve these equation using either analytical methods, which involve deriving formulas [Hello 1990, Vinet 2009], or numerical methods, which utilize computer computation. The choice between these methods depends on the complexity of the model.  The temperature response function of the mirror is determined by the power drive of the heating unit, and the thermal deformation response function {$f_{i,j} = f_i(x_j)$} along the sampling position $x_j$ is computed accordingly.  The target modulation shape $h_j=h(x_j)$ can be generated by 
\begin{equation}
h_{j} =\sum f_{j,i} p_{i}
\end{equation}
Where $p_{i}$ is the driving power of $i-th$ heat unit. To derive the inverse solution for this equation, the driving power $p_{i}$ can be determined using the non-linear least squares method implemented in SciPy, an open-source Python library. Fig. 1(b) and 1(c) illustrate thermal fields and thermal deformations modified through a simulation experiment using Finite Element Analysis (FEA). Considering disparities between the boundary conditions and system characteristics in the real system and the simulated model, the response coefficient of simulation data needs to be calibrated at the initial stage of the application. It should be noted that this system does not exactly require precise measurements of every system parameter, as the experiment that follows will demonstrate. 

The active optics verification test system is illustrated in Fig. 1(d)-(f). Following the simulation model,  heating elements are arranged on the top of an optical fused quartz glass mirror with a surface flatness of $\lambda$/4, as shown in Fig. 1(d)(e). The size of the fused quartz mirror is 205 mm in length, 50 mm in width, and 20 mm in thickness. The X-ray reflection area (3 mm * 200 mm) is located in the center portion along the tangential direction. There are 18 pairs of electrical heating elements (2.5 mm * 7.8 mm) with a 10mm period on both sides of the X-ray reflection zone. In order to prevent electrical conduction among the heating elements, a precise insulation gap of 0.2 mm is maintained between each unit. To simplify the fabrication process, a single mask is utilized to deposit both types of layers onto the substrate simultaneously, resulting in the formation of a 200 nm Cr coating. The Copper pad layer, depicted as the purple area in Fig 1(e), has a thickness of 2 $\mu$m. This thickness is chosen to ensure that the major thermal load is on the Cr resistor, resulting in uniform heating. Simulation calculations and physical intuition suggest that achieving greater surface heating deformation modulation performance can be achieved by symmetrically spreading the heating elements on both sides of the reflective coating and positioning them as close as feasible to the reflective coating. The MEMS (micro-electromechanical systems) process utilizes modern semiconductor microfabrication techniques to achieve precise micrometer-level accuracy. This results in cost-effective and consistent heating elements, each having a resistance of 13.3 ±0.1$\Omega$.

The heating elements are powered by six units of programmable DC power supply (Keithley 2230G1). The minimum division of a single heating element can approach 0.02 mW (current 1 mA) based on the measured resistance of the heating elements and the power supply precision. This thermal-deformable mirror ensures the reliable implementation of the first-principles design for two specific reasons: (1) precise model of heating source. the heating elements function by utilizing joule heating directly on the mirror surface. The structure of the mirror is simple, consisting only of the mirror body and coating. The manufacturing precision of the heating elements is at the micrometer level; (2) simple boundary conditions. The heat generated by the heating elements primarily transfers through the substrate via heat conduction and through convective heat exchange with the surrounding air, which implies that the boundary conditions for FEA simulations are explicitly defined. Hence, the thermal deformation characteristics (shape) of the entire system are deterministic due to its precise structure and distinct boundary types. Adjusting the computational parameters enables a precise estimation of actual operational conditions.

Finally, a pp-LTP (penta-prism long-trace profiler) shown in Fig.1 (f) was constructed and employed to measure the extremely small thermal deformation resulting from the heating elements on the mirror. The thermal-deformable mirror is housed in a small stainless steel chamber because high-precision measurements require a very steady environment. With the use of a scanning pentaprism inside the chamber and a lab-built optical head outside, the pp-LTP allows for accurate scanning measurements of the surface shape of large mirrors. The coating in the reflection area simultaneously serves as the measurement condition for the pp-LTP. When considering the practical uses of X-ray reflective mirrors, several coating materials can be selected according to specific criteria.
\begin{figure}
\subfigure{
\includegraphics[width=0.45\textwidth]{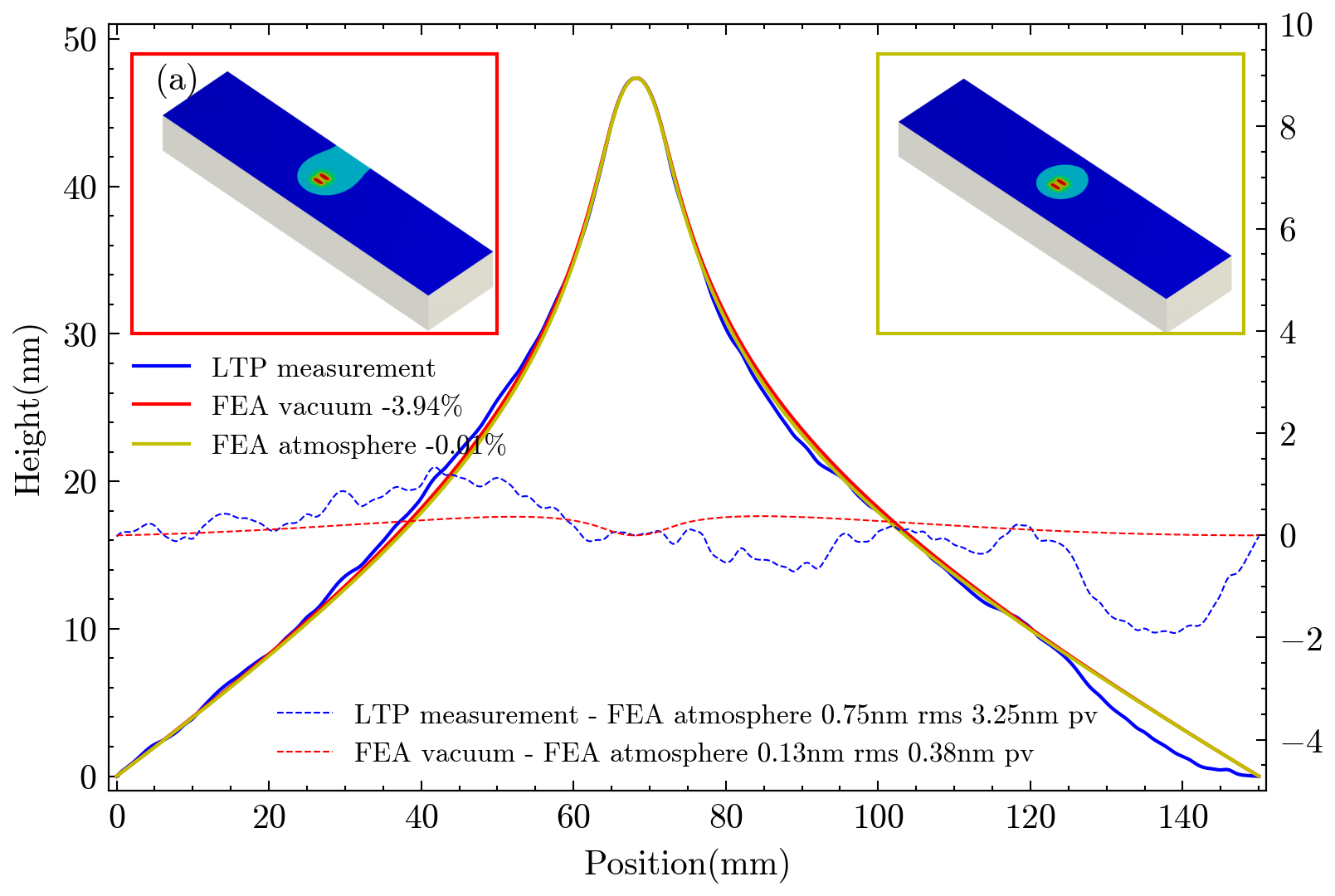}
\label{fig:subfig1}
}
\subfigure{
\includegraphics[width=0.45\textwidth]{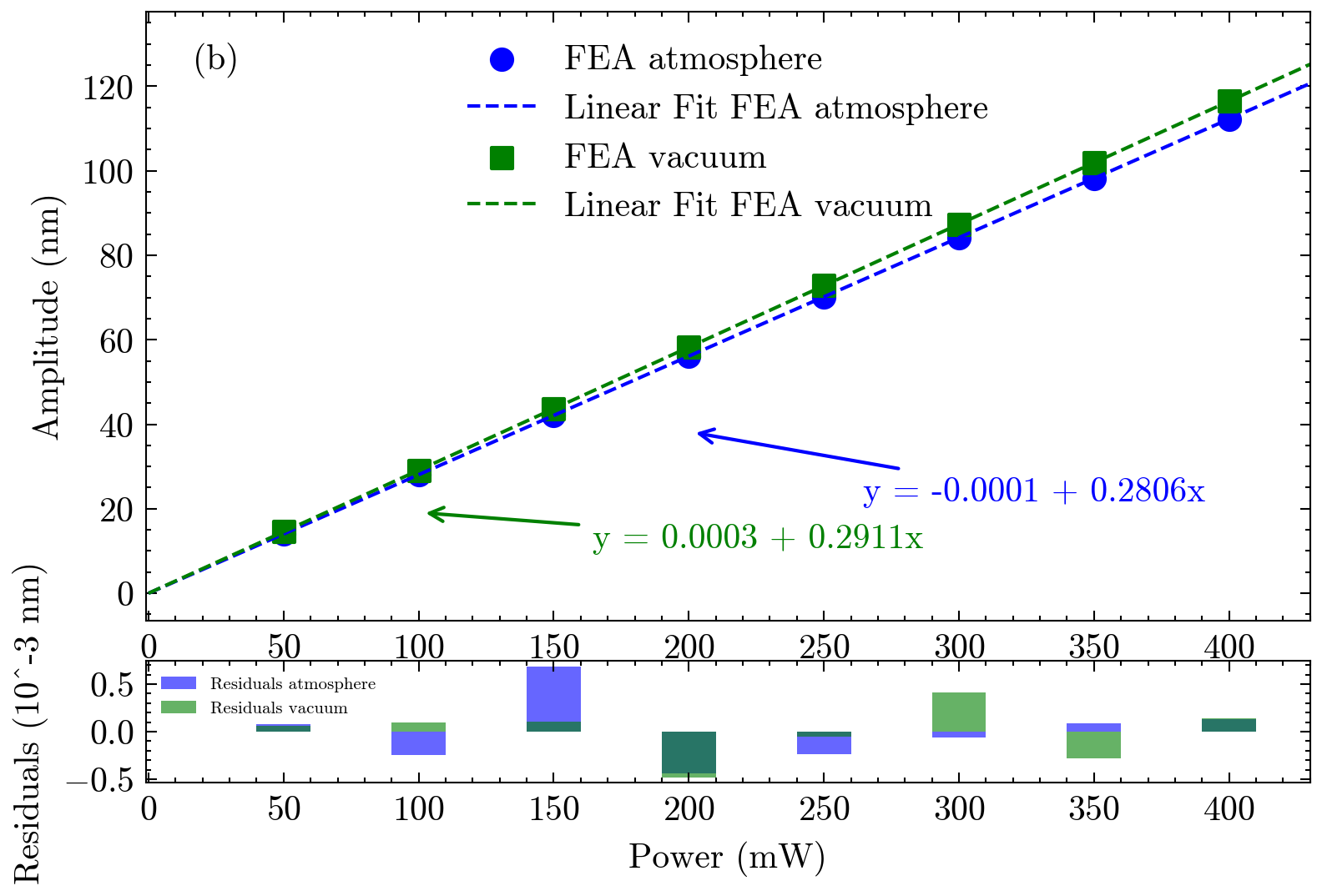}
\label{fig:subfig2}
}
\caption{Characterization the thermo-elastic deformation response, the 9th unit was selected. (a) compares the results of the first-principles calculation with the measured curve of the response function. The simulation conditions considered different boundary conditions: air convection and vacuum. Inset: 3D shape of the deformed mirror. The deformation amplitude determined by FEA is normalized to the measured shape by a factor of 3.94\% in vacuum and 0.01\% in air. The residual images (Y-axis 2) exhibit shape residual proportions of 0.75 nm rms and 0.13 nm rms, respectively. (b) shows the linear relationship between the thermal deformation response amplitude and the heating power, with a residual fitting at the $10^{-3}$ nm PV level.}
  \label{fig:2}
\end{figure}

The simulation and surface shape measurement results of the heating element deformation response are shown in Fig. 2. The calculation model involves applying a power of 0.2 W to the 9th pair of elements (see Supplementary Material 1.1). The thermal power contribution not contributing to the outward thermal expansion of the substrate surface is excluded, and the amplitude of the performance is scaled down (with a correction of 30\%). The test results indicate that, when suitably modifying a parameter, the deformation response of a single heating unit matches well with the FEA calculation results for the measured surface shape, exhibiting a residual of 0.75 nm rms(root mean square). Furthermore, we can see that the surface shape measurement data exhibits high-frequency oscillation characteristics, which is a common problem in surface shape metrology (due to high spatial frequency features or ambient measurement noise). In order to evaluate the influence of boundary conditions, the thermal deformation of the mirror is computed under vacuum conditions, employing side cooling and removing air convection. After normalization, it becomes apparent that the two-dimensional distributions exhibit dissimilarities. However, the deformations along the center line, which are particularly significant in X-ray applications, display an almost identical pattern. The difference between these deformations is 0.13 nm rms, observed at a deformation amplitude of 48 nm. To find out if the amplitude of the thermal deformation response is linear while taking into account the limits of measuring the surface shape accurately, a set of finite element simulation experiments was carried out to find the largest deformation response for various power drivers. The results, as depicted in Fig. 2(b), provide a linear property of 0.28 nm/mW for this heating component, accompanied by a fitting residual of about $10^{-3}$ nm PV. Based on FEA calculations, it can be proved that for a given mirror structure, deviations in material properties and fluctuations in heat transfer or thermal boundary conditions have minimal impact on the shape of thermal deformation at low levels of deformation. However, they have an effect on the magnitude of the thermal deformation response. The amplitude of the thermal deformation response can be precisely adjusted by calibration tests, similar to with other linear systems.

\begin{figure}[h]
  \subfigure{
    \includegraphics[width=0.45\textwidth]{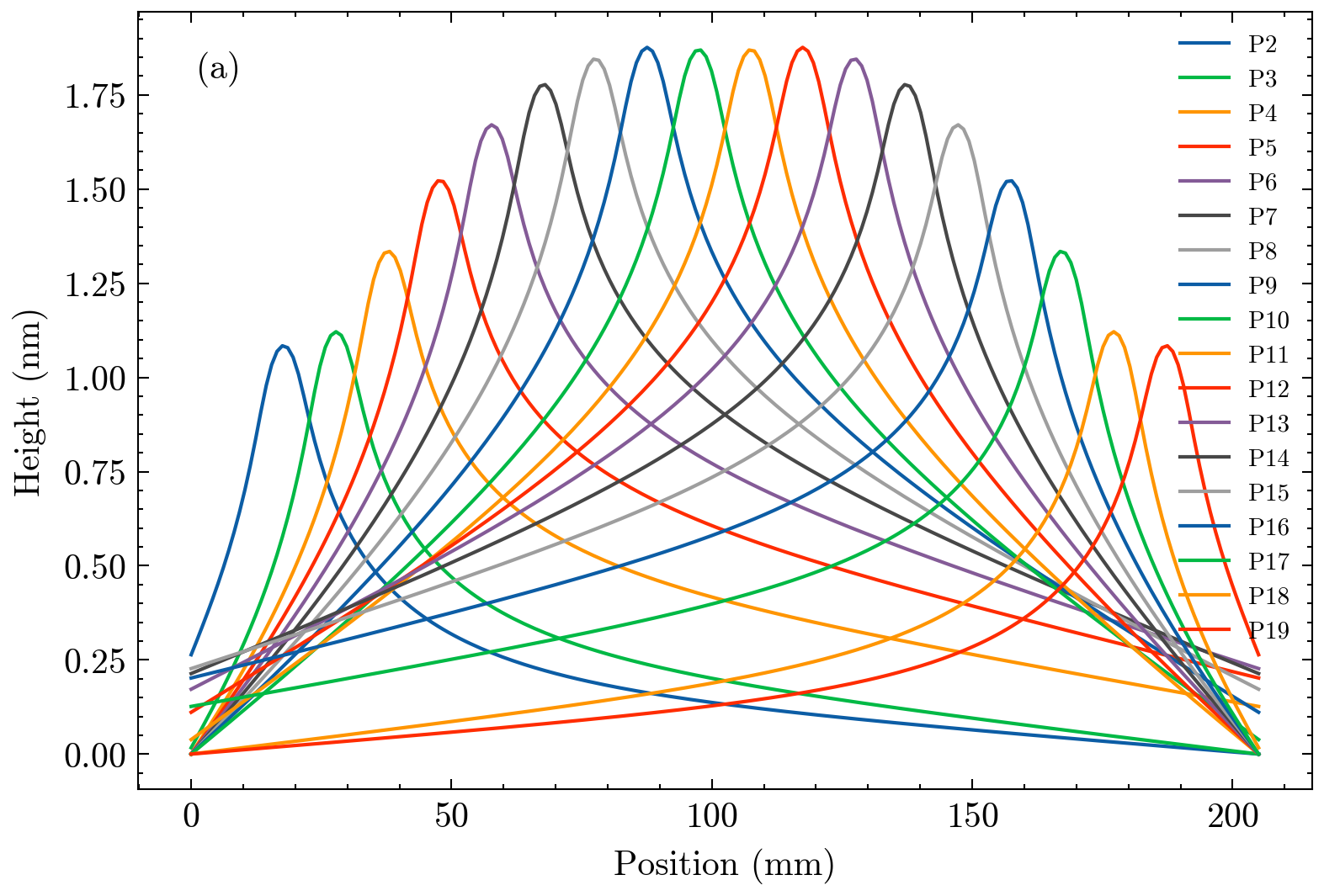}
    \label{fig:subfig1}
  }
  \subfigure{
    \includegraphics[width=0.45\textwidth]{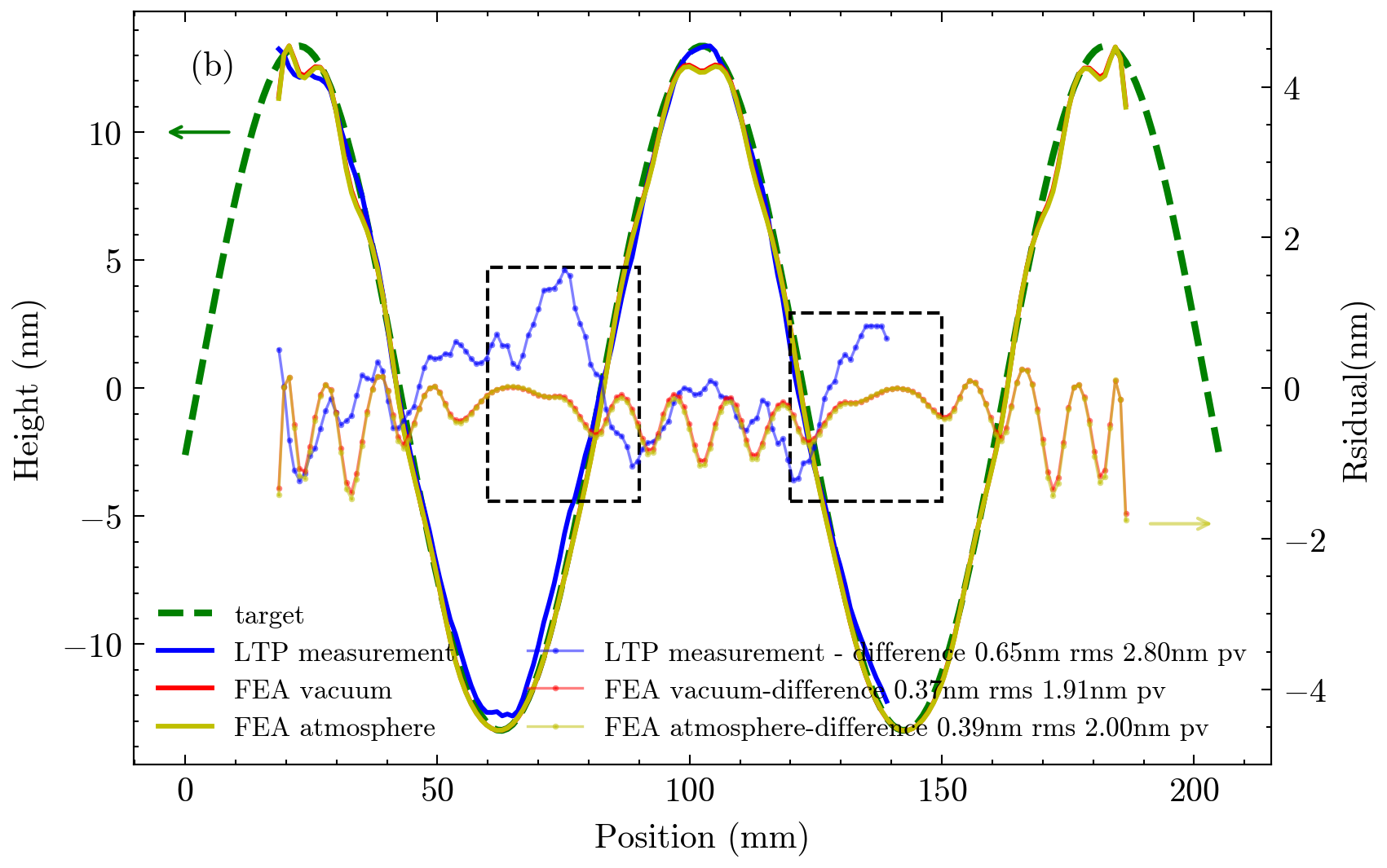}
    \label{fig:subfig3}
}
\caption{\label{fig:3} Results of the thermal deformation modulation experiment. (a) the thermal deformation response curve calculated using FEA under air conditions. (b) resulting modulation curve for a sinusoidal target with an amplitude-period of 80 mm, FEA simulation and measured results are presented with the boundary condition in the calculations being the mirror's bottom cooling. Theoretical calculations for vacuum condition are also given for comparison.}
\end{figure}

To characterize the shape modulation capability of the thermal-deformable mirror, we experimentally measured the modulated surface shape. The FEA and experimental results are shown in Fig. 3. Firstly, the response functions for each unit are calculated in FEA and depicted in Fig. 3(a), with a loading power of 0.2 W. Due to the wider heat dissipation path at the ends of the reflective mirror compared to the middle, the thermal deformation at these heating elements becomes smaller and asymmetric. Considering that the thermal elastic deformation decreases with increasing distance, modulation units at the edge region contribute less and pose a challenge to achieving modulation accuracy compared to the central region, introducing the edge effects of thermal deformation. The surface modulation target is configured as a sinusoidal deformation with a period of 80 mm. The pairs of heating elements are powered up in accordance with the computation to generate sinusoidal deformation. In order to reduce the influence of edge effects on the precision of modulation, we have shortened the merit-function optimization region (the range of $x_{j}$) in our experiment by 40 mm compared to the complete length of the deformable mirror. Note that all 18 pairs of heating units are involved in the optimization calculation. The results of FEA simulation experiments and PP-LTP surface shape measurement experiments are shown in Fig. 3(b), with a total electrical power of 1W for the 18 pairs of heating elements. Similar to the single-point response test, considering the differences between the physical system parameters and the simulation calculation system, the target deformation in the calculation is scaled to fit the measured surface shape. The final adjustment coefficient consistent with the measured sinusoidal shape is 70\%, with an amplitude of ±14 nm and modulation capability of 0.028 nm/mW. The surface results optimized based on finite element data and the surface detection results based on pp-LTP show differences from the target surface shape of 0.39 nm rms and 0.65 nm rms, respectively. The main contribution to the residual of the former comes from the discrete arrangement of a finite number of heating units and the corresponding sampling phase adaptation for the given sinusoidal wave deformation.  The optimization of the phase discrepancy between the heating element arrangement and the modulation target function is not addressed in this work.
The high-frequency features of the residuals in the measured data (circled in Fig. 3(b)) show that the disparity between the theoretical modulation residuals and the observed data is due to defects in the surface profiling system. With respect to the influence of boundary conditions, Fig. 3(b) concurrently displays the results under vacuum conditions in finite element simulations, akin to the computations shown in Fig. 2. In this case, the adjustment coefficient is 0.6, and the residual value of 0.37 nm rms is basically consistent with the test results in the atmosphere. This suggests that the changes in these boundary conditions solely impact the magnitude of deformation.

\begin{figure}[h]
  \subfigure{
    \includegraphics[width=0.45\textwidth]{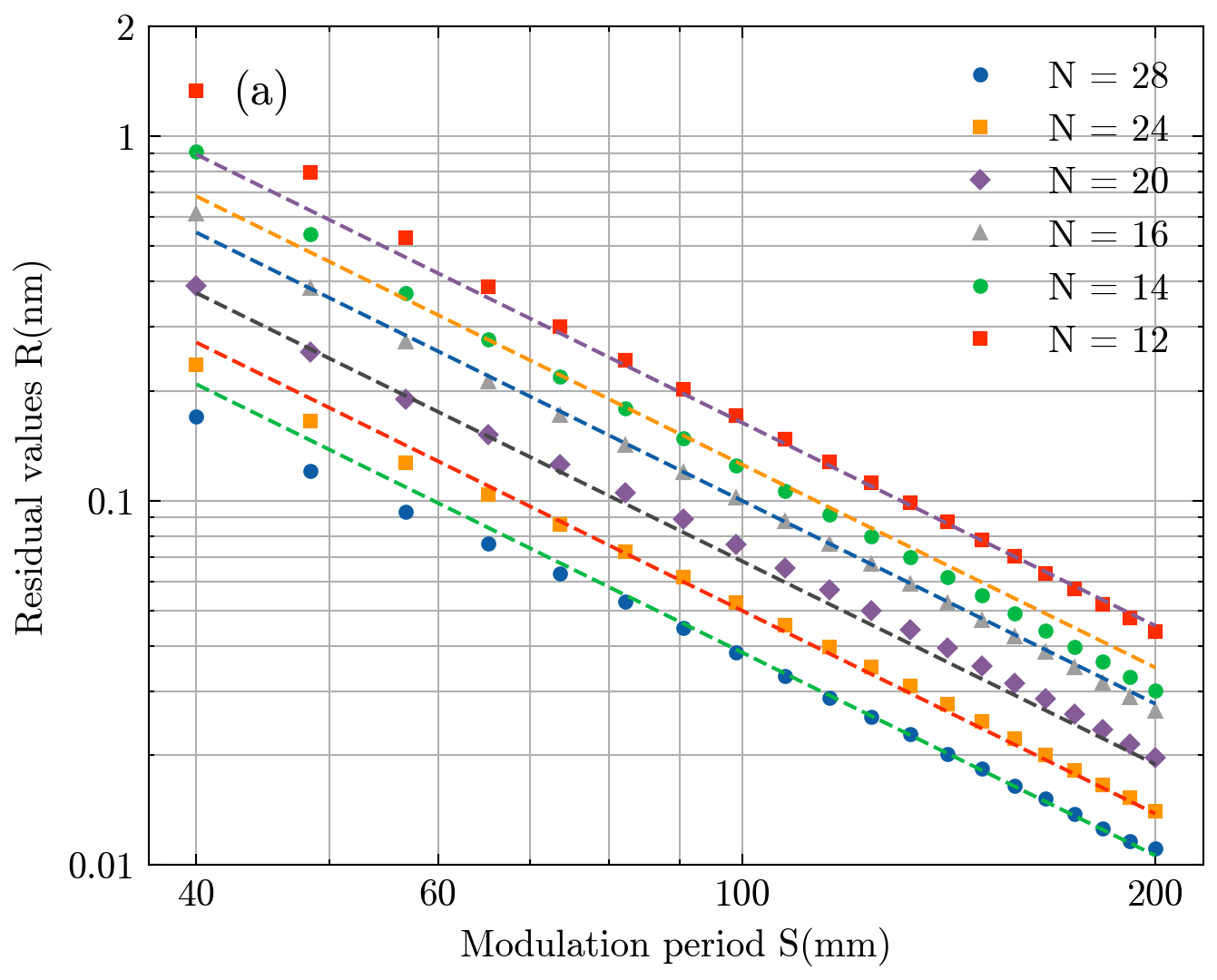}
    \label{fig:subfig4a}
  }
  \subfigure{
    \includegraphics[width=0.45\textwidth]{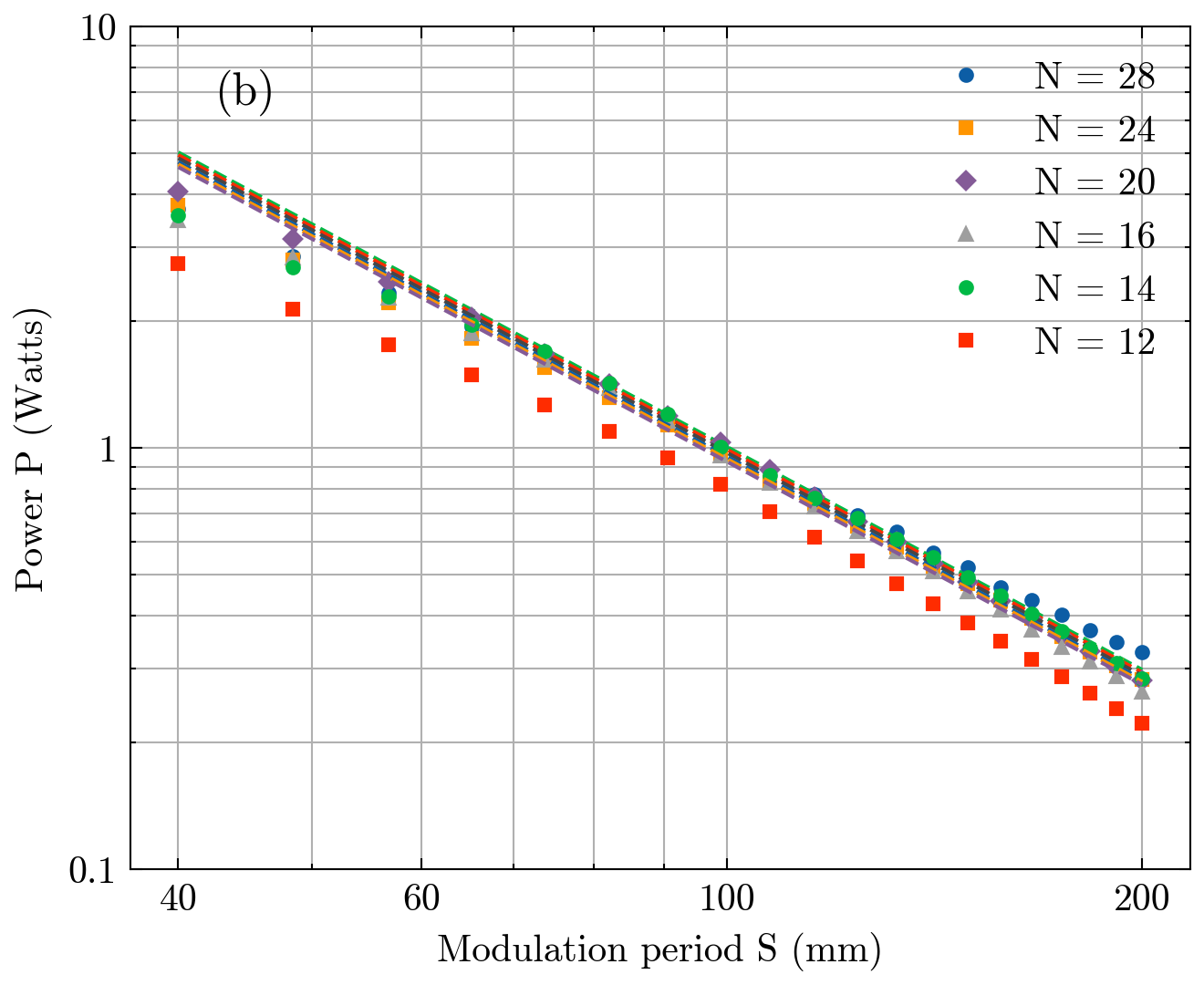}
    \label{fig:subfig4b}
  }
\caption{\label{fig:4} Thermal deformable mirror optimization taking into account the number of heating elements $N$ and the period of target sinusoidal shape $S$. (a) residual shape errors $R$; (b) total driving power $P$. The power function fit results are also plotted, with R-squared values of 0.987 and 0.995, respectively, for $log(P) = 7.78-1.76log(T)+0.09logN$ and $log(R) = 10.9-1.85log(T)-1.71logN$.}
\end{figure}

In order to ensure accuracy in a system, it is essential to allocate an acceptable error budget for each stage of the processes involved, such as fabrication, assembly, and correction. Optimizing the thermal deformation modulation to achieve optimum efficiency, particularly focusing on precision and power consumption, is significantly important for system design.  The studies conducted indicate the significance of limiting the magnitude of deformations in a certain heating unit design when we consider the linear response characteristics of thermal elastic deformation. Another approach entails adjusting the dimension of heating units by considering the distribution of the desired deformation in the spatial frequency domain. In the following study, we investigated the impact of modifying the size and power of heating elements on the ability to form sinusoidal target shapes with different periods but the same amplitude. The residual shape error required as well as the total heating power were assessed. This analysis is depicted in Fig. 4. The amplitude of the sinusoidal target functions with varied periods is set to 5 nm, taking into account the typical polishing precision of an X-ray mirror. Fig. 4(a) shows that the accuracy of modulation improves progressively as either the number of heating elements N or modulation period S increases. It is found that the negative power function of $R = 54176/(S^{1.85}N^{1.71})$ can be used to fit the curves with R-squared value of 0.995. The fit error becomes large for high-frequency shape modulation (i.e. small $S$). The power required and minimum residual value for various sinusoidal modulations $S$ are shown in Fig. 4(b). The result shows that higher-frequency surface modulation needs a lot of heating power and leaves a big shape error, which is in line with how general linear systems respond. The low power requirement in the case of N = 12 can be explained by the low modulation accuracy capability. The power function of $P = 2392N^{0.09}/(S^{1.76})$ with R-squared value of 0.987 is also a good fit for each curve when the heating element number $N$ is larger than 14. Note that the overall heating power required for surface modulation in the entire system should be controlled considering the challenge of the cooling technique. To achieve a match between the number of heating elements (power investment) and modulation accuracy, we can fit the minimum number of elements required to achieve the corrected accuracy for the target. By the fitting function in Fig.4 (a), the residual error is approximately proportional to $(SN)^{-1.8}=(S*L/S_{heating})^{-1.8}$, which means that the requirement of the heating elements within a single modulation period $S/S_{heating}$ is more fundamental. For instance, to achieve a low-frequency surface correction residual of 0.1 nm rms, $S/S_{heating}$ should be larger than 7.7 in this system where $L$ = 200 mm. Overall, the final mirror design should consider the characteristics of the desired shapes. For a particular shape and the design of the heating components, raising the heating power can result in large deformation but reduced modulation accuracy.

In summary, we present an attempt to build an X-ray active optical system through the application of first-principles computations. We have successfully demonstrated the effectiveness of our design approach in thermal elastic deformation mirrors, where heating elements are directly and accurately manufactured on the mirror's surface.  The agreement between the simulations and the real modulation performance of this technology is confirmed through optical testing on the long trace profiler. In comparison with other active optical systems that have limitations because of complex, indirect correction mechanisms and the requirement of accurate surface shape metrology, the deterministic modulation capability of X-ray active optics system shows promise in achieving precise sub-nanometer-level form modulation at a low cost. It is crucial to emphasize the universal value of this precision instrument design methodology based on first principles. In an age of computer science, this approach enables the efficient development of precision instruments and efficiently manages costs. In particular, the first-principles-based active optics system design is a paradigm shift, representing a technique freed from the limitations of conventional machining and polishing. New-generation synchrotron radiation facilities can benefit from this X-ray modulation technology that enables precise wavefront control over diffraction-limited focusing and beam shaping.

\begin{acknowledgments}
This work is supported by the National Science Foundation of China (No. 11505212, 11875059). We thank Prof. Zhongliang Li, Prof. Le Kang, Prof. Hongxin Luo for their experimental help and useful discussions.
\end{acknowledgments}

\nocite{*}

\bibliography{apssamp}

\end{document}